\documentclass[aps,prl,twocolumn,groupedaddress]{revtex4}
\usepackage{graphicx}
\usepackage{amsmath}
\usepackage{natbib}

\newcommand{\pd}{{h}}        
\newcommand{\kp}{{k_0}}             
\newcommand{\kf}{{k_+}}             
\newcommand{\kb}{{k_-}}             
\newcommand{\kbp}{{k_{\mathrm bp}}}         
\newcommand{\ku}{{k_u}}      
\newcommand{\Ld}{{\ell}}            
\newcommand{\dU}{{\Delta U}}        
\newcommand{\Sc}{{\cal S}_{cis}}    
\newcommand{\St}{{\cal S}_{trans}}  
\newcommand{\dG}{{\Delta G}}        
\newcommand{\cF}{{\cal F}}          

\begin{document}

\title{Coupled dynamics of RNA folding and nanopore translocation}

\author{Ralf Bundschuh$^1$} 
\author{Ulrich Gerland$^2$}

\affiliation{
$^1$Department of Physics, The Ohio State University, Columbus, Ohio \\ 
$^2$Arnold-Sommerfeld Center for Theoretical Physics and Center for Nanoscience (CeNS), LMU M\"unchen, Theresienstrasse 37, 80333 M\"unchen, Germany}

\date{\today}

\begin{abstract} 
The translocation of structured RNA or DNA molecules through narrow pores necessitates the opening of all base pairs. Here, we study the interplay between the dynamics of translocation and base-pairing theoretically, using kinetic Monte Carlo simulations and analytical methods. We find that 
the transient formation of basepairs that do not occur in the ground state can significantly speed up translocation.
\end{abstract}

\pacs{???}
\maketitle

\begin{figure}[b]
\includegraphics[width=8cm]{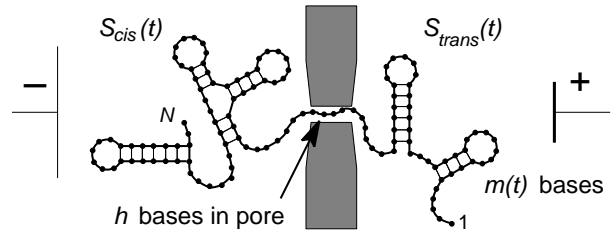}
\caption{Translocation of a structured RNA molecule through a narrow pore, which allows single but not double strands to pass. The driving force is an applied voltage, which acts on the negatively charged backbone of the RNA. An appropriate reaction coordinate is the number of bases $m(t)$ that have reached the {\it trans} side. The dynamics of $m(t)$ is coupled to the dynamics of the base-pairing patterns $\Sc(t)$ and $\St(t)$ on the {\it cis} and {\it trans} sides.}
\label{fig1}
\end{figure}

The dynamics of base-pairing in DNA and RNA molecules plays an important role in many biological processes ranging from DNA replication to RNA folding \cite{Alberts:02}. Often, the dynamics is coupled to other kinetic processes. For instance, RNA folds as it is synthesized, while it unfolds and refolds as it passes through the ribosome during translation. A series of recent single-molecule experiments \cite{experiments} used electric fields to drive RNA and DNA through tiny pores, which let single but not double strands pass. The case of unstructured, e.g. homopolymeric molecules, is particularly well characterized, with many theoretical studies complementing the experiments \cite{homopolymer_translocation,Lubensky:99,anomalous_translocation}. In contrast, the effect of base-pairing on translocation, see Fig.~\ref{fig1}, is only beginning to be explored, with a few existing experiments \cite{Vercoutere:01,Sauer-Budge:03,Mathe:04} and first steps towards a theoretical description \cite{Sauer-Budge:03,Gerland:04}. In this letter, we specifically examine the interplay between base-pairing and translocation dynamics. 

Already homopolymer translocation displays rich dynamics. Experimentally, one can obtain the full distribution $p(\tau)$ of the times $\tau$ for the translocation of a molecule from the {\it cis} to the {\it trans} side of the pore. Hence, the goal of theoretical descriptions is to determine $p(\tau)$ or its moments as a function of system parameters such as the chain length $N$, the voltage drop across the pore, and temperature. Most treatments reduce the dynamics to a one-dimensional reaction coordinate, $m$, which measures how many bases have reached the {\it trans} side \cite{homopolymer_translocation,Lubensky:99}. This approach is justified only when the polymer degrees of freedom equilibrate rapidly compared to the timescale of translocation. It fails for long chains, where the internal polymer dynamics and friction with the solvent limit the translocation speed \cite{anomalous_translocation}. However, most experiments so far were performed with short chains where the friction between the polymer and the pore dominates the translocation dynamics \cite{experiments,Lubensky:99}. In a minimal model for these experiments the index $m$ is increased with a forward rate $\kf$ and decreased with a backward rate $\kb$. When expressed in terms of a continuous reaction coordinate $0<x<N$, the drift-diffusion equation 
\begin{equation} 
\partial_{t} P(x,t) = D\,\partial_{x}^2 P(x,t) - v\,\partial_{x} P(x,t) 
\label{Eq_Drift_Diffusion}
\end{equation}
describes the time evolution of the probability distribution for $x$, where $D$ is an effective diffusion constant and the drift velocity $v$ is induced by the applied voltage. The ratio $\Ld=D/v$ defines a diffusive lengthscale, below which the drift $v$ is negligible compared to the Brownian motion \cite{Lubensky:99}. From (\ref{Eq_Drift_Diffusion}), one obtains $p(\tau)$ as the probability current into the absorbing boundary at $x\!=\!N$. 

Base-pairing interactions affect the translocation dynamics significantly by introducing sequence- and structure-dependent kinetic barriers \cite{Vercoutere:01,Sauer-Budge:03,Mathe:04}. Previous theoretical work \cite{Gerland:04} argues that these barriers can be exploited to measure the secondary structure of an RNA molecule, if its sequence is already known. However, the interplay of the translocation and base-pairing dynamics also poses new physical questions, which will be important for the design of applications. For instance, one expects two limiting cases for the dynamics: {\em slow translocation}, during which the base-pairing pattern remains equilibrated at almost all times and on both sides of the pore, and {\em fast translocation}, where an essentially frozen base-pairing pattern on the {\it cis} side is unzipped as the molecule passes through the pore. Ref.~\cite{Gerland:04} considers only the latter limit. Here, we are interested in the crossover from the slow to the fast translocation limit and address the following questions: (i) Can the physics of the crossover be described in simple terms? (ii) What is the nature and size of the effective free energy barriers for translocation? (iii) How is the distribution $p(\tau)$ affected by the presence of secondary structure ? 

Whether an experiment is closer to the slow or fast translocation limit can depend on the sequence/structure of the molecule under study, the bias voltage, and the friction coefficient between the polymer and the pore. The latter determines the rate $\kp$ for translocation by a single (unpaired) base at zero bias, which we use as a control parameter for the crossover. Microscopically, $\kp$ is determined by the diameter of the pore and its surface chemistry, both of which are in principle tunable, either by mutation in the case of protein pores or through the fabrication process in the case of solid-state pores \cite{Li:01,Storm:03}. We will see below that the pore depth is also a relevant parameter for the questions raised above.

{\it Model.---}
We focus on the case of RNA, even though our results apply equally to single-stranded DNA. Since kinetic barriers slow the translocation of structured RNA, we expect the description by the reaction coordinate $m$ to be valid over an even wider regime of lengths $N$ than for unstructured RNA. However, for structured RNA, the dynamics of $m(t)$ is coupled to the dynamics of the RNA base-pairing patterns $\Sc(t)$, $\St(t)$ on both sides of the pore. In our model, illustrated in Fig.~\ref{fig1}, only unpaired bases can enter the pore. We describe the dynamics of $\Sc(t)$ and $\St(t)$ using Monte Carlo kinetics with three elementary moves: opening of a pair, closing, or a shift in the binding partner of a base \cite{Flamm:00}. We use a single rate $\kbp$ for all moves that are energetically favorable, and the rate $\kbp\,e^{-\dG}$, if the move increases the Gibbs free energy by $\dG$ (all energies are in units of the thermal energy $k_{B}T$). While this assumption does certainly not hold on the microscopic level, we expect that it will not affect the qualitative features of the dynamics on long timescales. Indeed, a recent model of force-induced RNA unfolding \cite{Harlepp:03} has shown a remarkably good agreement with experiment using similar assumptions.


To calculate the Gibbs free energies we use a simplified energy model, which allows for CG, AU, and GU basepairs with a binding energy of 2, 1, and 0, respectively. The free energy cost for any loop (interior, bulge, multiloop, or hairpin) of size $\ell$ is $3\,\nu\ln(\ell)$, where $\nu\approx 0.6$ is the Flory exponent. We exclude hairpin loops of size $\ell<3$, which cannot occur in RNA structures due to steric constraints. The applied voltage drops primarily directly across the pore, leading to an energy gain $\dU$ when a monomer traverses the pore. Thus, the ratio of the forward and backward rates is given by the Boltzmann factor $e^{\dU}$. We assume the symmetric choice $k_{\pm}=\kp\,e^{\pm\dU/2}$ corresponding to a centered transition state along the microscopic reaction pathway for translocation by one monomer. We denote by $\pd$ the number of bases that fit inside the pore and assume these cannot base-pair at all. While translocation consists of an entrance stage followed by a passage stage, we focus only on the passage dynamics. Hence, we choose $m(0)=0$ as initial condition and stop the simulation when $m(\tau)=N$, whereas we do not allow the RNA to exit on the {\it cis} side.

\begin{figure}[t]
\includegraphics[width=8.6cm]{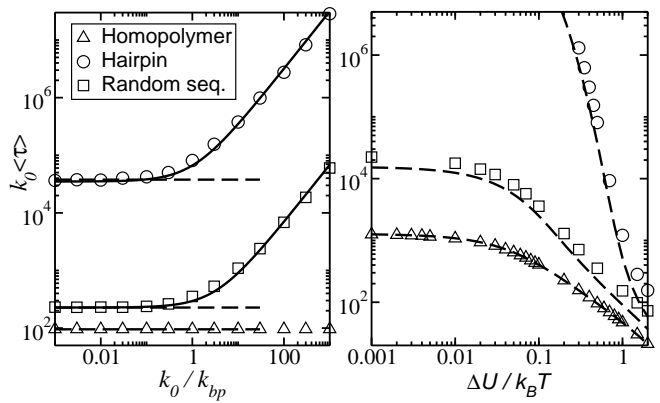}
\caption{Mean translocation time $\langle\tau\rangle$ for RNAs with the same length $N=50$, but different structures (we plot $\langle\tau\rangle$ in units of $\kp^{-1}$, which is $O(\mu s)$ in typical experiments \cite{experiments}) . Symbols represent simulation data, while dashed lines show the adiabatic limit (\ref{EQmeantau}), and solid lines the expression (\ref{EQ2RW}) for the two random walker model.
(a) Crossover from slow to fast translocation at a fixed bias $\dU=0.5\,k_{B}T$. Typical basepairing rates are $\kbp\!\sim\!1\,\mu s^{-1}$ \cite{Craig:71}, so that typical experiments would fall into the  crossover region. 
(b) Dependence on the bias at a fixed ratio $\kp/\kbp=1$.
}
\label{fig2}
\end{figure}

{\it Mean translocation times.---}
To explore the effect of base-pairing on the translocation dynamics, we use three different sequences with identical length $N=50$; one 
well-defined structure (a hairpin with 23 random base pairs and a loop of 4 bases), one random sequence, and an unstructured homopolymer for comparison. Fig.~\ref{fig2} shows their mean translocation times $\langle\tau\rangle$ for different combinations of the parameters $\kp$, $\kbp$, and $\dU$ (here we have used an idealized pore with $\pd=1$). The average is taken over 1000 independent simulations, and $\langle\tau\rangle$ is plotted in units of the 'hopping time' $\kp^{-1}$ for an unpaired base across the pore at zero bias.

Fig.~\ref{fig2}a displays the transition from the slow to the fast translocation limit by varying the ratio of the two characteristic timescales in our model, $\kp/\kbp$, at constant bias $\dU\!=\!0.5$. For the homopolymer, the base-pairing timescale is irrelevant and hence $\kp\langle\tau\rangle$ is constant. For the structured RNAs, the curves also approach a constant limit at small $\kp/\kbp$, indicating that translocation is sufficiently slow to allow equilibration of the basepairing pattern whenever the RNA translocates by a base. Note that this limit value of $\kp\langle\tau\rangle$ depends strongly on the sequence/structure of the RNA. In the other extreme, $\kp\gg\kbp$, base-pair opening in front of the pore is the rate limiting process. The crossover between these limits occurs at $\kp\sim\kbp$. We will now seek a quantitative understanding of these simulation results.

{\it Slow translocation limit.---}
In the limit $\kp\ll\kbp$, the translocation dynamics $m(t)$ becomes independent of the base-pairing kinetics. The effect of base-pairing then amounts to reducing the translocation rates $\kf$, $\kb$ by factors equal to the probability that the base entering the pore is unpaired. Such equilibrium properties of the base-pairing interaction can be expressed through the partition function $Z_{i,j}$, the total statistical weight of all secondary structures for the substrand from base $i$ to base $j$ of the RNA sequence. For instance, the probability for the $j$-th base on the {\it trans} side to be unbound is $Z_{1,j-1}/Z_{1,j}$. Hence, the translocation dynamics $m(t)$ becomes a 1D random walk with site-dependent hopping rates $\kb(m)=\kb Z_{1,m-1}/Z_{1,m}$ and $\kf(m)=\kf Z_{m+\pd+2,N}/Z_{m+\pd+1,N}$. Using the mean first passage time formalism \cite{Gardiner:83}, we then find the mean translocation time 
\begin{equation}
	\label{EQmeantau}
  \langle\tau\rangle = \frac{1}{\kf}
  \sum_{j=m(0)}^{N-1}\,\sum_{k=0}^{j} 
  e^{\cF_{h+1}(j)-\cF_{h}(k)-(j-k)\dU} \;,
\end{equation}
where $\cF_{\pd}(m)=-\ln(Z_{1,m}Z_{m+\pd+1,N}/Z_{1,N})$ is the free energy cost of placing a free RNA from solution into position $m$ of a pore with depth $\pd$. We obtain the partition functions $Z_{i,j}$ by generalizing the well-known recursion relations of \cite{McCaskill:90} to a logarithmic multiloop cost \cite{Note1}, in order to derive the exact adiabatic limit of our model. 

\begin{figure}[t]
\includegraphics[width=\columnwidth]{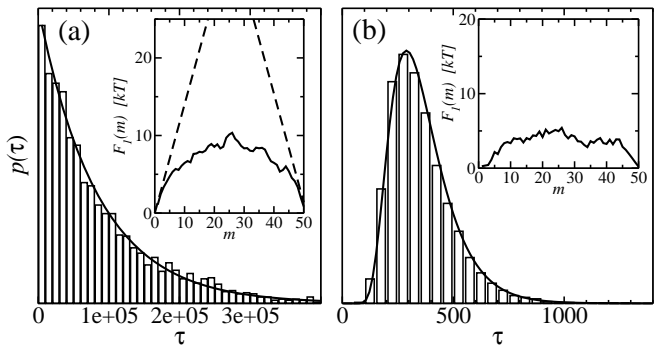}
\caption{Translocation time distributions for (a) the RNA hairpin and (b) the random sequence at $\dU=0.5$ and $\kp=\kbp$. The histograms show simulation data, whereas the solid line in (a) is an exponential fit and in (b) a fit to the drift-diffusion form \cite{Lubensky:99}. The insets show the free energy landscapes $\cF_{1}(m)$ of the adiabatic limit (\ref{EQmeantau}), see main text for details.}
\label{fig3}
\end{figure}

In Fig.~\ref{fig2}a, the dashed lines show that our kinetic simulations indeed converge to the adiabatic limit (\ref{EQmeantau}) for small $\kp/\kbp$. However, besides providing a check on our simulations, Eq.~(\ref{EQmeantau}) also helps to understand the nature of the relevant free energy barriers. For example, the hairpin structure has a total binding free energy of $\approx 33\,k_{B}T$. Yet, its translocation time compared to that of the homopolymer is not more than 1000 times longer (at $\dU=0.5\,k_{B}T$ and $\kp<\kbp$). The applied voltage lowers the free energy barrier only about $10\,k_{B}T$ and does not suffice to explain the comparatively small effect of the stable structure on the translocation time. Inspection of the free energy landscape $\cF_{1}(m)$ of the hairpin, shown in the inset of Fig.~\ref{fig3}a, reveals that the formation of non-native base-pairs on both sides of the pore lowers $\cF_{1}(m)$ considerably in the central part, and explains the effect (for comparison, the landscape for only the native hairpin structure is shown with a dashed line). Hence, the major barrier for the translocation of structured molecules is not unfolding of the complete structure, but partial unfolding of the structure until the formation of non-native basepairs speeds up translocation.

{\it Crossover to fast translocation.---}
In Fig.~\ref{fig2}a, the crossover between slow and fast transition occurs in the range $0.1<\kp/\kbp<10$ for both structured RNA sequences. Incidentally, typical experimental values for the base-pairing and translocation rates are comparable with $\kbp\!\sim\!\kp\!\sim\!1\,\mu s^{-1}$ \cite{experiments,Craig:71}. Hence it is important to characterize the translocation behavior in the transition region. 

Note that while the higher stability of the hairpin yields a longer translocation time, the shape of the prominent bend in the curves of Fig.~\ref{fig2}a is independent of the sequence as long as it is structured. This observation encourages us to develop a simplistic model: The dominant effect giving rise to the prominent bend is a kinetic competition between the opening/closing dynamics of the base pair next to the {\it cis} side of the pore, and the position of the pore with respect to the RNA, which is driven into the structure by the external bias. This process resembles two impenetrable random walkers biased to run into each other. Walker `P' representing the pore has a hopping rate $\kf$ towards walker `R' representing the RNA and a rate $\kb$ away from `R'. Walker `R', in turn, moves towards `P' at the rate $\kbp$ for base pair closing, and away at an unbinding rate $\ku$ that is reduced by the stability of the base pair. If none of these rates depends on the positions of `R' and `P', one finds that the center of mass position of the walkers drifts with a velocity $v=(\ku\kf-\kbp\kb)/(\kbp+\kf)$. Assuming this drift dominates over diffusion, $\langle\tau\rangle$ takes the simple form
\begin{equation}
\label{EQ2RW}
\langle\tau\rangle=N/v=\frac{N}{\kp}\left(a+b\frac{\kp}{\kbp}\right)
\end{equation}
with dimensionless constants $a$ and $b$. The solid lines in Fig.~\ref{fig2}a show Eq.~(\ref{EQ2RW}), where we determined $a$ from the slow translocation limit (\ref{EQmeantau}) and used $b$ as fitting parameter. The agreement with the simulation data suggests that the shape of the curves is indeed largely determined by a kinetic competition between translocation and the local base-pairing dynamics in the vicinity of the pore. The effects of large-scale rearrangements in the base-pairing pattern during translocation are too subtle to be discernible in Fig.~\ref{fig2}a. However, reformation of base-pairs on the {\it trans} side, which we have ignored to obtain Eq.~(\ref{EQ2RW}), strongly influences the parameter $b$.

{\it Voltage dependence.---}
In Fig.~\ref{fig2}b the mean translocation time of our three test sequences is plotted against the bias $\dU$, where symbols represent simulation data at $\kbp=\kp$ and dashed lines the adiabatic limit (\ref{EQmeantau}). For the homopolymer, the adiabatic limit is exact and displays the well-known crossover from a diffusion-dominated $\langle\tau\rangle\sim N^2/\kp$ at $\dU\ll1/N$, to a drift-dominated $\langle\tau\rangle\sim N/\kf$ at $\dU\gg1/N$. For the structured RNAs, the mean translocation time also levels off at $\dU\sim1/N$, since the molecule ``feels'' the bias only if its total free energy differs by at least one $k_{B}T$ between the two sides of the pore. In this case, however, the simulation data lies somewhat above the adiabatic limit (\ref{EQmeantau}), consistent with the fact that $\kbp=\kp$ is within the crossover regime of Fig.~\ref{fig2}a. Clearly, as we leave the adiabatic limit, the effective kinetic barriers for translocation increase and (\ref{EQmeantau}) gives only a lower bound on the mean translocation time. As $\dU$ is increased to $\sim k_{B}T$ (right edge of the plot), the system is more strongly perturbed and the deviation from the adiabatic limit becomes more significant.

{\it Translocation time distributions.---}
In Fig.~\ref{fig2}b, the translocation dynamics of the homopolymer is drift-dominated for $\dU>1/N$. Is this equally true for the translocation dynamics of the structured RNAs? To address this question, we examine the 
full distribution $p(\tau)$, which displays more detailed signatures of the translocation dynamics than the average $\langle\tau\rangle$ and is routinely measured experimentally. The histograms in Figs.~\ref{fig3}a and \ref{fig3}b show $p(\tau)$ for (a) the hairpin and (b) the random sequence, from simulations at $\kp=\kbp$ and $\dU=0.5\gg1/N$. We observe that the distributions are markedly different: For the hairpin, $p(\tau)$ follows an exponential law (solid line) characteristic for thermally activated crossing of a single free energy barrier. In contrast, $p(\tau)$ for the random sequence has the shape characteristic for the drift-diffusion process (\ref{Eq_Drift_Diffusion}), which is appropriate for {\em unstructured} polymers (the solid line in Fig.~\ref{fig3}b is a fit with the associated mean first passage time distribution \cite{Lubensky:99}). 

Indeed, the free energy landscape $\cF_{1}(m)$ of the random sequence is remarkably flat (see inset of Fig.~\ref{fig3}b). This is due to the fact that the random sequence can accommodate many dissimilar structures with similar free energies: when the pore brakes up part of the structure of the random sequence, a (non-native) structure can reform on the {\it trans} side, which is almost equally stable. Hence, the translocation dynamics of the random sequence is qualitatively similar to that of the homopolymer, only with a reduced diffusion coefficient $D$ and drift velocity $v$ due to the roughness of the free energy landscape.

{\it Conclusions.---}
Compared to the thermodynamics, the dynamics of base-pairing is far less understood. Translocation experiments are emerging as a valuable tool to probe these dynamics on the single-molecule level. Here, we have studied the interplay between translocation and base-pairing within a simplified, but explicit theoretical model. In the slow translocation limit, the translocation dynamics of our model can be understood analytically, which is useful to interpret our simulations even outside the slow translocation regime. We find that the formation of non-native basepairs can significantly speed up translocation, rendering the effective kinetic barriers much smaller than might be naively expected. Furthermore, we showed that the crossover between slow and fast translocation can be described by a simple phenomenological model. In the future, nanopores might be used to sort RNAs according to their structures.

We gratefully acknowledge discussions with R. Neher and funding by the Petroleum Research Fund of the ACS through Grant No.~42555-G9 (to RB), by the Ohio Supercomputer Center (to RB), and by an Emmy Noether Grant of the DFG (to UG).


\begin{thebibliography}{22}

\bibitem{Alberts:02}
B.~Alberts {\it et al.}, \emph{Molecular Biology of the Cell} (Garland, 2002).

\bibitem{experiments}
J. Kasianowicz, E. Brandin, D. Branton, and D. Deamer, Proc. Natl. Acad. Sci. USA \textbf{93}, 13770 (1996).
A.~Meller, J. Phys.: Condens. Matter \textbf{15}, R581 (2003).
A.J. Storm, J.H. Chen, H.W. Zandbergen, and C. Dekker, Phys. Rev. E \textbf{71}, 051903 (2005).

\bibitem{homopolymer_translocation}
W.~Sung and P.J. Park, Phys. Rev. Lett. \textbf{77}, 783 (1996).
M.~Muthukumar, J. Chem. Phys. \textbf{111}, 10371 (1999).
E.~Slonkina and A. Kolomeisky, J. Chem. Phys. \textbf{118}, 7112 (2003).

\bibitem{Lubensky:99}
D. Lubensky and D. Nelson, Biophys. J. \textbf{77}, 1824 (1999).

\bibitem{anomalous_translocation}
J.~Chuang, Y.~Kantor, and M. Kardar, Phys. Rev. E \textbf{65}, 011802 (2002).
A.~Milchev, K.~Binder, and A.~Bhattacharya, J. Chem. Phys. \textbf{121}, 6042 (2004).

\bibitem{Vercoutere:01}
W.~Vercoutere {\it et al.}, Nat. Biotechnol. \textbf{19}, 248 (2001).

\bibitem{Sauer-Budge:03}
A. Sauer-Budge, J. Nyamwanda, D. Lubensky, and D.~Branton, Phys. Rev. Lett. \textbf{90}, 238101 (2003).

\bibitem{Mathe:04}
J.~Mathe {\it et al.}, Biophys. J. \textbf{87}, 3205 (2004).

\bibitem{Gerland:04}
U.~Gerland, R.~Bundschuh, and T.~Hwa, Phys. Biol. \textbf{1}, 19 (2004).

\bibitem{Li:01}
J.~Li {\it et al.}, Nature \textbf{412}, 166 (2001).

\bibitem{Storm:03}
A.J. Storm {\it et al.}, Nature Materials \textbf{2}, 537 (2003).

\bibitem{Flamm:00}
C.~Flamm, W.~Fontana, I. Hofacker, and P.~Schuster, RNA \textbf{6}, 325 (2000).

\bibitem{Harlepp:03}
S.~Harlepp {\it et al.}, Eur. Phys. J. E \textbf{12}, 605 (2003).

\bibitem{Gardiner:83}
C. Gardiner, \emph{Handbook of Stochastic Methods} (Springer, Berlin, 1983).

\bibitem{McCaskill:90}
J.~McCaskill, Biopolymers \textbf{29}, 1105 (1990).

\bibitem{Note1}
Whereas \cite{McCaskill:90} linearizes the free energy of multi-loops, we retain the logarithmic form derived from polymer theory. The detailed calculation will be published elsewhere. Briefly, we split $Z_{i,j}$ into a contribution $B_{i,j}$ from structures including the base-pair $(i,j)$, and contributions $U_{i,j}^{\ell}$ without this base-pair, $Z_{i,j}=B_{i,j} + \sum_{\ell=0}^{j-i+1} U_{i,j}^{\ell}$, where $\ell$ is the number of exterior unbound bases. We denote by $Q_{j,j-1}^{i,i+1}$ the Boltzmann factor for the stacking of the basepairs $(i,j)$ and $(i+1,j-1)$. For $B_{i,j}$ we then obtain $B_{i,j} = Q_{j,j-1}^{i,i+1}\,B_{i+1,j-1} + \sum_{\ell} \ell^{-3\nu} \, U_{i+1,j-1}^{\ell}$, if the bases $i$ and $j$ match, and zero otherwise. Similarly, we find $U_{i,j}^{\ell} = U_{i+1,j}^{\ell-1} + \sum_{k} B_{i,k} (U_{k+1,j}^{\ell} + \delta_{\ell,0}\,B_{k,j}) + \delta_{\ell,1} B_{i+1,j}$. With these relations, and the initial conditions $B_{i,j}=0$ and $U_{i,j}^{\ell}=\delta_{\ell,j-i+1}$ for $j-i<4$, the partition functions are calculated in $O(N^4)$ time --- slower than the $O(N^3)$ time of~\cite{McCaskill:90}, but  feasible for sequences up to 1000 bases.

\bibitem{Craig:71}
M.E. Craig, D.M. Crothers, and P. Doty, J. Mol. Biol. \textbf{62}, 383 (1971).

\end{thebibliography}
\end{document}